\documentclass[a4paper,11pt]{article}
\pdfoutput=1 

\usepackage{jcappub} 

\usepackage[T1]{fontenc} 

\usepackage{multirow} 

\title{\boldmath 
A Neural-Network based estimator to search for 
primordial non-Gaussianity in Planck CMB maps
}

\author[a,c]{C. P. Novaes,}
\author[a]{A. Bernui,}
\author[b]{I. S. Ferreira}
\author[c]{and C. A. Wuensche}

\affiliation[a]{Observat\'orio Nacional,\\Rua General Jos\'e Cristino 77, S\~ao Crist\'ov\~ao, 20921-400, Rio de Janeiro, RJ, Brazil}
\affiliation[b]{Instituto de F\'{\i}sica, Universidade de Bras\'{\i}lia, Campus Universit\'ario Darcy Ribeiro,\\Asa Norte, 70919-970, Bras\'{\i}lia, DF, Brazil}
\affiliation[c]{Divis\~ao de Astrof\'isica, Instituto Nacional de Pesquisas Espaciais,\\Av. dos Astronautas 1758, S\~ao Jos\'e dos Campos, 12227-010, SP, Brazil}

\emailAdd{camilapnovaes@gmail.com}
\emailAdd{bernui@on.br}
\emailAdd{ivan@fis.unb.br}
\emailAdd{ca.wuensche@inpe.br}

\abstract{
We present an upgraded combined estimator, based on Minkowski Functionals and Neural Networks, with excellent performance in detecting primordial non-Gaussianity in simulated maps that also contain a weighted mixture of Galactic contaminations, besides real pixel's noise from Planck cosmic microwave background radiation data. 
We rigorously test the ef\/ficiency of our estimator considering several plausible scenarios for residual non-Gaussianities in the foreground-cleaned Planck maps, with the intuition to optimize the training procedure of the Neural Network to discriminate between contaminations with primordial and secondary non-Gaussian signatures. 
We look for constraints of primordial local non-Gaussianity at large angular scales in the foreground-cleaned Planck maps. 
For the $\mathtt{SMICA}$ map we found ${f}_{\rm \,NL} = 33 \pm 23$, at $1\sigma$ confidence level, in excellent agreement with the WMAP-9yr and Planck results.
In addition, for the other three Planck maps we obtain similar constraints with values in the interval 
${f}_{\rm \,NL} \in [33, 41]$, concomitant with the fact that these maps manifest distinct features in 
reported analyses, like having different pixel's noise intensities.
}

\begin{document}
\maketitle
\flushbottom

\section{Introduction} \label{sec:introduction}

In the standard cosmological scenario, the \textit{cosmic inflation} \citep{1981/guth} is responsible for the origin of structures, like galaxies, clusters, and voids, we observe today. 
Inflation not only explains the large-scale homogeneity and isotropy of the universe, but also provides a mechanism to produce primordial fluctuations in cosmic microwave background (CMB) radiation which follow a nearly Gaussian statistics, and where deviations from it characterises dif\/ferent classes of inflationary models~\citep{2009/abramo, 2009a/kawasaki, 2009b/kawasaki, 2009/komatsu, 2010/komatsu, 2010/chen}. Severe constraints for Gaussian deviations found in the latest Planck analysis strongly supports the simplest inflationary model based on a single minimally-coupled scalar field~\citep{2014/planck-XXII}. However, as already discussed in~\cite{2014/planck-XXIV, 2014/planck-XXII, 2014/planck-XXIII}, there are potential non-Gaussian sources that could leave imprints in Planck maps, these include a scale-dependent non-Gaussianity (NG) and/or large-scale anisotropy. In addition to these non-Gaussian cosmological signals, contaminations could be caused either by residual foregrounds or by instrumental systematics. 

Non-Gaussian contributions appear mixed in Planck maps, with diverse phenomena contributing with their own signature, this makes necessary the use of different estimators. Various statistical methods are being used in the analyses of CMB data in order to constrain the distinct types of NG, to measure their intensity and angular scale dependence~\citep{2003/komatsu, 2012/WMAP9a, 2004/bartolo, 2010/bartolo, 2014/bernui, 2015/bernui, 2012/ducout, 2013/modest, 2014/novaes, 2011/raeth, 2014/planck-XXIII, 2014/planck-XXIV}. Moreover, all methods designed to analyse precise CMB data are vulnerable to additional dif\/ficulties in detecting tiny primordial NG, that is, the contamination by several secondary non-Gaussian signals, such as galactic and extragalactic foregrounds, besides systematic ef\/fects. For this, it is crucial to test estimators to discriminate between these two type of signals, making sure not be attributing a primordial origin to a secondary non-Gaussian signal~\citep{2010/bartolo, 2010/komatsu, 2014/planck-XXIV}.

In \citep{2014/novaes} we have presented a statistical estimator which proved to be quite sensitive to detect tiny primordial NG of local type, endowed in synthetic Gaussian CMB maps, besides its capability to dif\/ferentiates between primary and secondary (inhomogeneous noise) NGs. This method used the Minkowski Functionals (MFs) \citep{1903/minkowski} calculated from a set of Monte Carlo (MC) CMB maps as input data for the training of an Artificial Neural Network (NN)~\citep{1999/haykin}, that classifies each synthetic map of a second dataset according to the presence and level of NG. To test our estimator in real CMB maps we have used the four CMB foreground-cleaned, maps released by the Planck in March of 2013, namely Spectral Matching Independent Component Analysis ($\mathtt{SMICA}$), Needlet Internal Linear Combination ($\mathtt{NILC}$), Spectral Estimation Via Expectation Maximization ($\mathtt{SEVEM}$), and the combined approach termed $\mathtt{Commander-Ruler}$ \citep{2014/planck-XII}. Together with each of these four CMB maps were also released its realistic pixel's noise map, that we have used to contaminate the MC CMB maps with secondary NG. In that work our analyses showed that, in the circumstances in which the method have been tested, the NN is not able to recognise the full pattern of NG present in the Planck CMB maps. These results indicated that there are non-Gaussian contributions in Planck data that are not present in MC simulations, resulting in the imprecision of the method. 

In this work we address this problem by exploring possible residuals candidates left in foreground-cleaned maps. We do this examination by extending previous analyses to include more sources of NG, besides the primordial one. This is made by including several mixtures of small weighted contributions of primordial and non-primordial NG in the current MC CMB maps in order to suitable train the NN. The Galactic non-Gaussian contributions considered are the residual foreground emissions, which include synchrotron, free-free and dust emissions, the most important Galactic emissions in the frequency range we are dealing with here (70, 217 GHz), beyond the inhomogeneous noise already considered in our previous work. The extragalactic non-Gaussian contributions, including for example the Sunyaev-Zel'dovich (SZ) and lensing ef\/fects, besides point sources (secondary CMB anisotropies), are not relevant for the scales considered. In addition, the Planck Galactic mask used here, termed U73, includes the point-sources mask~\cite{2010/bartolo, 2012/aluri, 2013a/munshi, 2008/aghanim, 2014/planck-XII}. In this way, our main goal is to test the ef\/ficiency of our estimator to constrain primordial NG in the presence of non-Gaussian foreground contaminations from diverse Galactic emissions. Specifically, this work extends and complements our previous analyses in three ways: 
\begin{itemize}
\item[(i)] 
We test the ef\/fectiveness of our estimator in simulated maps endowed with a residual Galactic foreground component, which is a weighted mixture of dust and free-free emissions, besides a tiny primordial non-Gaussian signal. The contribution of these contaminating emissions comes from two frequency bands, that is, 70, and 217 GHz. Moreover, these mixtures consider two levels of contamination for each emission, 0.1\% and 10\%, in an ef\/fort to reproduce plausible levels of residuals in Planck maps.
\item[(ii)] 
The MC CMB maps have been contaminated by noise maps derived from real pixel's noise (released by the Planck collaboration) and standard deviation per pixel maps released in association with the $\mathtt{SMICA}$, $\mathtt{NILC}$, and $\mathtt{SEVEM}$ Planck CMB maps. For the $\mathtt{Commander-Ruler}$ Planck map it was also released the real noise map. It was generated a set of ten noise maps related to each one of these four foreground-cleaned Planck maps, in order to obtain dif\/ferent combinations between CMB anisotropies and noise.
\item[(iii)] 
We rigorously test the performance of our method considering several scenarios of NG possible present in Planck maps, with the intuition to optimize the training procedure of the NN to discriminate between contaminations with primordial and secondary non-Gaussian signatures. 
\end{itemize}

In the next section we describe all the procedures to obtain the synthetic data employed here. 
We continue this paper summarising, in section~\ref{sec:estimator}, the main features of the tools composing our combined estimator, in addition to detailing how it operates. In section~\ref{sec:application} we show the results of applying our estimator to synthetic and Planck CMB data. Finally, in section~\ref{sec:conclusion}, we present our concluding remarks.

\section{Data description} \label{sec:data}

The development of the current work requires a large amount of synthetic data, as will be detailed in following sections. These data are quite dif\/ferent from those used in our previous work, since now we are interested in studying the ef\/ficiency of our estimator in the presence of a mixture of residual foreground contaminations in the MC CMB maps. Therefore, our datasets contain combinations of simulated components, namely: MC CMB maps, inhomogeneous noise and Galactic emissions (synchrotron, free-free and dust emissions). The last one will be used in a set of frequencies and diverse levels of contaminations, working as a residual foreground emission. 

All the maps used here are produced with ${\rm \,N}_{\mbox{\small side}} = 512$, using the HEALPix (Hierarchical Equal Area iso-Latitude Pixelization) pixelization grid~\citep{2005/gorski}. The details of the production of each component are as follows. In all our analyses we use the U73 cut-sky mask.

\subsection{The Monte Carlo CMB maps} \label{subsec:cmb} 

We generate the MC CMB maps from a set of coef\/ficients {$a_{\ell \, m}$} (with maximum multipole moment of $\ell_{max} = 500$) derived from a combination of the multipole expansion coef\/ficients $\{ a^{\mbox{\footnotesize G}}_{\ell \, m} \}$ and $\{ a^{\mbox{\footnotesize NG}}_{\ell \, m} \}$ corresponding to CMB Gaussian and non-Gaussian (of local type) maps, respectively. A set of each kind of these spherical harmonics coef\/ficients, 1000 linear and 1000 non-linear, were produced and publicly available by \cite{2009/elsner}\footnote{http://planck.mpa-garching.mpg.de/cmb/fnl-simulations/}, and we combine them as follows:

\begin{equation} \label{alms_comb}
\, a_{\ell \, m} = a^{\mbox{\footnotesize G}}_{\ell \, m} 
+ f_{\rm \,NL} \, a^{\mbox{\footnotesize NG}}_{\ell \, m} \, ,
\end{equation}

\noindent which were normalized by the Planck best fit power spectra \citep{2014/planck-I}, rescaling $a^{\mbox{\footnotesize G}}_{\ell \, m}$ by the ratio of the square root of the power spectra and $a^{\mbox{\footnotesize NG}}_{\ell \, m}$ directly by the ratio of the power spectra. The scalar dimensionless parameter $f_{\rm \,NL}$ is commonly used to describe the leading-order of the NG. Then the linear combination given in the Equation \ref{alms_comb} permits to derive CMB synthetic maps with an arbitrary level of NG defined by any real value of $f_{\rm \,NL}$. We also used a HEALPix beam window function with FWHM = 5' for the construction of these maps\footnote{The HEALPix pixel window at ${\rm \,N}_{\mbox{\small side}} = 512$ is also included.}, since the angular resolution of Planck maps corresponds to a Gaussian beam of this size. Moreover, we also used 1000 CMB maps obtained from the original ones after a clockwise rotation of 90 degrees perpendicularly to the Galactic plane. 
This procedure give us a total of 2000 CMB base maps, from which we construct the 
training and test simulated maps (using the chosen set of $f_{\rm \,NL}$ values). 

In this work we use three ranges of $f_{\rm \,NL}$ values to include a primary non-Gaussian signal in the simulations. In order to test our estimator in MC CMB maps containing the degree of contamination defined by the recent constraints found by {\it Planck}~\citep{2014/planck-XXIV}, that is, $f_{\rm \,NL} = 38 \,\pm\, 18$ (1$\sigma$ confidence level), we choose the levels of contamination in the following ranges: $f_{\rm \,NL} \,=\, [-20,20], [20,60]$ and $[60,100]$. Each interval is composed by an uniform distribution of $f_{\rm \,NL}$ values centred in 0, 40, and 80, respectively. Notice that these features dif\/fers of those considered in Ref. \citep{2014/novaes}, where we assumed a Gaussian distribution of $f_{\rm NL}$ values to compose each class.

\subsection{Residual Galactic foreground contamination} \label{subsec:galaxy} 

We have tested our estimator in MC CMB maps contaminated by residual foreground emission in two frequency bands, chosen to be 70 and 217 GHz. In this case, the most important Galactic foreground contributions are the synchrotron, free-free and dust emissions. The dust maps on these frequencies are estimated performing an interpolation \cite{2008/de-oliveira-costa}, pixel by pixel, from a set of dust maps which covers a wide range of frequencies, maps publicly available as part of the Planck\footnote{http://irsa.ipac.caltech.edu/data/Planck/release\_1/all-sky-maps/previews/COM\_CompMap\_dust-commrul\_2048\_R1.00/index.html} and WMAP-9yr\footnote{http://lambda.gsfc.nasa.gov/product/map/dr5/mem\_maps\_get.cfm} data releases \citep{2014/planck-XII,2013/bennett}. A similar approach is followed to derive the synchrotron and free-free maps, this time performing an extrapolation to the frequencies we are interested in, using the available  maps of these foreground diffuse emissions. It is worth mentioning that WMAP-9yr foreground maps used here were derived using  the Maximum Entropy Method (MEM) (see, e.g., Ref. \citep{2013/bennett}).

As mentioned, the Galactic foreground contamination was included to the MC CMB maps as a residual signal in an attempt to imitate the possible contents of a data map. In other words, we add the simulated foreground map to the MC CMB map after it was weighted by a percentage factor, that we will call here the \textit{weight}. For definiteness, we consider two weight values for the foreground contamination: 0.1\% and 10\% of the foreground map. In this way, we can also test the sensitivity of the estimator to dif\/ferent levels of contamination concerning this kind of signal.

A last detail to be considered refers to the right way to include this kind of contamination. 
In~\cite{2014/planck-XII} \textit{Planck team} uses the analysis of the FFP6 simulations (Full Focal Plane 6, see \cite{2013/planck-ES} for details), exactly the same performed upon the real data, to provide important informations, or previsions, about the residual contamination expected for the four released Planck CMB maps. This type of contamination takes the following form on the simulation results: (1) the contamination on the $\mathtt{Commander-Ruler}$ is an under-subtracted free-free emission; (2) for the $\mathtt{NILC}$ and $\mathtt{SEVEM}$ it seems to be an over-subtracted thermal dust emission; and (3) for the $\mathtt{SMICA}$ there is an under-subtracted thermal dust emission. Therefore, the contamination is performed by adding a weighted foreground map to the MC CMB maps, when aiming to mimic the $\mathtt{SMICA}$ and $\mathtt{Commander-Ruler}$ residual maps, and by subtracting it from these maps if necessary to mimic the $\mathtt{SEVEM}$ and $\mathtt{NILC}$ residual maps. 

\subsection{Inhomogeneous Planck-like noise} \label{subsec:noise}

We simulate four types of noise, namely, the $\mathtt{Commander-Ruler}$, $\mathtt{SMICA}$, $\mathtt{NILC}$ and $\mathtt{SEVEM}$-like noise, using the corresponding standard deviation map. However, only the standard deviation map associated to the $\mathtt{Commander-Ruler}$ CMB map has been released by Planck. For the other three cases, we estimate them from the provided noise maps, taking the $\sigma_{noise}$ in each pixel $p$ as being the standard deviation of the noise values from a set of pixels in a disk around $p$. Multiplying the standard deviation maps, pixel-by-pixel, by a normal distribution with zero mean and unitary standard deviation we obtain the Planck-like noise maps.

We generate a total of 10 maps of each kind of noise in such a way to provide dif\/ferent combinations between the MC CMB temperature fluctuations and noise.

\subsection{Summary of the simulated data}

The purpose here is to produce MC CMB maps contaminated by secondary NG derived from 
Galactic foregrounds and inhomogeneous noise trying to follow, in the best possible way, the 
expected contamination to the Planck CMB maps due to such signals. 
Table \ref{tab:data} summarises the combinations of contaminants we include in the MC CMB maps 
in order to produce all the synthetic data used to construct and test our estimator. 
The reference map, in the first column, corresponds to the maps whose Galactic foreground and 
noise contaminants are being mimicked.

\begin{table}[h]
\centering
\begin{tabular}{| c | c c c |}
\hline
Reference map     & Noise                     &  Frequency (GHz)   & Weight (\%)  \\
\hline
$\mathtt{SMICA}$                   & $\mathtt{SMICA}$-like            &  70, 217 & 0.1, 10 \\
$\mathtt{NILC}$                      & $\mathtt{NILC}$-like               &       217         & 10    \\
$\mathtt{SEVEM}$                  & $\mathtt{SEVEM}$-like          &       217         & 10       \\ 
$\mathtt{Commander-Ruler}$   & $\mathtt{Commander-Ruler}$-like    &       70$^a$      & 10    \\
\hline
\multicolumn{4}{p{13.5cm}}{$^a${\footnotesize Between the two frequencies considered here, the 70 GHz frequency is the most free-free contaminated, directly related to the expected residual contamination of the Planck $\mathtt{Commander-Ruler}$ map (see section \ref{subsec:galaxy}).}} 
\end{tabular}
\caption{Simulated datasets.} \label{tab:data}
\end{table}

\section{The combined estimator} \label{sec:estimator}

Aiming to statistically analyse CMB data searching for a possible deviation from Gaussianity, we have been working on a high-performance estimator combining two tools: the Minkowski Functionals and Neural Networks. The first one, introduced on cosmological studies by \cite{1994/mecke}, is largely used for statistical analysis of the two-dimensional CMB field. Since it can measure morphological properties of fluctuation fields, it of\/fers a test of the Gaussian nature of the CMB temperature f\/luctuations data \citep{1999/novikov, 2003/komatsu, 2004/eriksen, 2006/naselsky_book, 2012/hikage, 2013/modest, 2013b/munshi}. As they are sensitive to weak and arbitrary non-Gaussian signals, e.g. small $f_{\rm \,NL}$ contaminations of dif\/ferent types, MFs are a complementary tool to optimal NG estimators based on the bispectrum calculations. 

Recently, the {\it Planck Collaboration} performed successful validation tests of the MF estimator with three sets of simulated CMB maps: Gaussian full-sky maps, full-sky non-Gaussian maps with noise, and non-Gaussian maps with noise and mask~\citep{2014/planck-XXIV}. Afterward, they used the MFs to quantify primordial NG in the foreground-cleaned {\it Planck} maps~\citep{2014/planck-XII}, where some instrumental ef\/fects and known non-Gaussian contributions, like lensing, were taken into account in the analyses using realistic lensed and unlensed simulations of {\it Planck} data~\citep{2014/planck-XXIV}. The constraints on local NG obtained are quite robust to Galactic residuals and are consistent to those from the bispectrum-based estimators. Moreover, these results, $f_{\rm \,NL} = 38 \pm 18$ for large angular scales, are basically equal to those obtained using WMAP-9yr data, that is $f_{\rm \,NL} = 37.2 \pm 19.9$~\citep{2012/WMAP9a, 2014/planck-XXIV}. 

In this work we do not use MFs in the standard form, that is, to directly quantify NG in the map, instead we use it just to reveal the non-Gaussian imprints present in the maps, signatures that are then systematically recognized by a well-trained NN. For this reason the MFs are employed as the first step to develop our estimator: first we apply the MFs to a set of synthetic maps; these parameters, with the intrinsic statistical properties associated to these maps, are then used as input for exhaustive analysis of the NN, a computational tool for identifying patters in a dataset. After this whole procedure the estimator is ready to be applied in dif\/ferent CMB maps, synthetic (to validate its performance) and real maps, allowing its classification about their level of NG. It is important to emphasize that the estimator we are dealing with here is the same addressed in our previous work~\citep{2014/novaes}, but this time we can interpret the NN's output in a way to directly estimate the $f_{\rm \,NL}$ parameter. 

In the next subsections we summarize the main concepts of each tool and how they are combined to 
obtain the estimator (see~\cite{2014/novaes} for details), in addition to explaining the new approach to addressing the NN's output.

\subsection{Minkowski Functionals} \label{subsec:mf}

All the morphological properties of a $d$-dimensional space can be described using $d+1$ MFs~\citep{1903/minkowski}. In the case of a CMB map, a 2-dimensional temperature field defined on the sphere, $\Delta T = \Delta T(\theta,\phi)$, with zero mean and variance $\sigma^2$, this tool provides a test of non-Gaussian features by assessing the properties of connected regions in the map. Given a sky path ${\cal P}$ of the pixelized CMB sphere ${\cal S}^2$, an {\em excursion set} of amplitude $\nu_t$ is defined as the set of pixels in ${\cal P}$ where the temperature field exceeds the threshold $\nu_t$, that is, it is the set of pixels with coordinates $(\theta,\phi) \in {\cal P}$ such that $\Delta T(\theta,\phi) / \sigma \equiv \nu > \nu_t$. 

In a two-dimensional case, for a region $R_i \subset {\cal S}^2$ with amplitude $\nu_t$ the partial MFs calculated just in $R_i$ are: $a_i$, the Area of the $R_i$ region, $l_i$, the Perimeter (contour length) of this Area, and $n_i$, the number of holes in this Area. The global MFs are obtained calculating these quantities for all the connected regions with height $\nu > \nu_t$. Then, the total Area $A(\nu)$, Perimeter $L(\nu)$ and Genus $G(\nu)$ are~\citep{1999/novikov, 2003/komatsu, 2006/naselsky_book, 2012/ducout}

\begin{eqnarray} \label{funcionais}
A(\nu) &=& \frac{1}{4 \pi} \int_{\Sigma} d\Omega = \sum a_i \, , \\
L(\nu) &=& \frac{1}{4 \pi} \frac{1}{4} \int_{\partial\Sigma} dl = \sum l_i \, , \\
G(\nu) &=& \frac{1}{4 \pi} \frac{1}{2 \pi} \int_{\partial\Sigma} \kappa dl = \sum g_i = 
N_{hot} - N_{cold} \, ,
\end{eqnarray}

\noindent
where $\Sigma$ is the set of regions with $\nu > \nu_t$, $\partial \Sigma$ is the boundary of $\Sigma$, and, $d \Omega$ and $dl$ are the elements of solid angle and line, respectively. In the Genus definition, the quantity $\kappa$ is the geodesic curvature (for more details see, e.g.,~\cite{2012/ducout}). This last functional can also be calculated as the dif\/ference between the number of regions with $\nu > \nu_t$ (number of hot spots, $N_{hot}$) and regions with $\nu < \nu_t$ (number of cold spots, $N_{cold}$). As defined, the MF are calculated from a given threshold $\nu_t$.

\subsection{Neural Networks} \label{subsec:nn}

The NNs are computational techniques inspired in the neural structure of intelligent organisms (animal brains), which acquires knowledge through learning \citep{1999/haykin}. A NN is composed by a large number of processing units (also called artificial neuron or nodes), configured to perform a specific action, like pattern recognition and data classification. 

\begin{figure}
\centering
\includegraphics[scale=0.7]{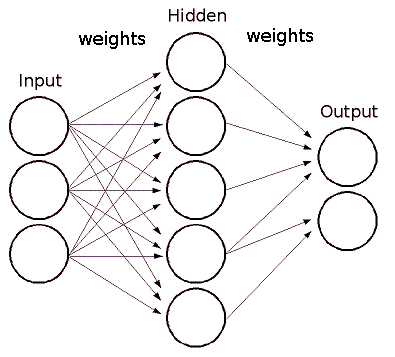}
\caption{Multilayer Perceptron.}
\label{NN}
\end{figure}

Aiming to emulate the behaviour of the biological brain, the simplest and most popular model for a NN for classification of patterns is the \textit{Perceptron} \citep{1943/mcculloch_pitts,1957/rosenblatt}, consisting of a single neuron. A generalization of this model is the \textit{Multilayer Perceptron} \citep{1986/rumelhart, 1998/zhang_patuwo, 1999/haykin}, consisting of a set of units (neurons) comprising each layer, from which the signal propagates through the NN. These NNs are usually composed by an input layer, an output layer and one or more intermediate (or hidden) layers, as schematically depicted in Figure~\ref{NN}. The layers are interconnected through synaptic weights ($w_{ki}$) that relates the $i$-th input signal ($x_i$) to the $k$-th neuron, producing weighted inputs. Mathematically it is possible to represent a neuron $k$ by 

\begin{equation}
u_k = \sum_{i=1}^{n} w_{ki} ~.~ x_i + b_k,
\end{equation}

\noindent where $n$ is the number of input signals to the $k$-th neuron. $b_k$ is the weighted bias input, an external parameter of the artificial neuron $k$, which can be accounted for by adding a new synapse, with input signal fixed at $x_0 = + 1$ and weight $w_{k0} = b_k$~\citep{1999/haykin}.

The output signal of the $k$-th neuron ($y_k$) is generated through an activation function $\varphi_k$, which limits the amplitude of the output of a neuron 

\begin{equation}
y_k = \varphi_k(u_k) \, .
\end{equation}

\noindent The most commonly used activation function are non-linear functions, like sigmoid and hyperbolic tangent, that can simulate more precisely the neuron behaviour in order to better emulate a real NN. 

The suitable values for the {\em weights} and {\em bias} are achieved by using training (or learning) algorithms. The most popular training algorithm is the \textit{backpropagation} \citep{2000/basheer}, where the input signal feeds the first layer, propagating as input for the next layer, and so on until the last layer. At the end of this process a value for each neuron of the output layer, together with its corresponding error, is calculated. These values are returned to the input layer where the synaptic weights are recalculated initiating another \textit{iteration} of the training process. This procedure aims to determine a suitable value for the weights of the network, and is repeated until the error value drops below a given threshold value. The architecture of the NN, like number of neurons, number of hidden layers and the specific training algorithm, can be defined according to the problem the user wants to solve. Here we used a backpropagation algorithm for the NN training with just one hidden layer, and 260 neurons, a number that has been shown to be suitable for the 3 classes we are analysing here.

\subsection{Minkowski Functionals as input for Neural Networks analysis} \label{subsec:mf_nn}

The calculations of the MFs used here are performed using the algorithm developed by~\cite{2012/ducout} and~\cite{2012/gay}. This code calculates four quantities, namely $V_0 = A(\nu)$, $V_1 = L(\nu)$, $V_2 = G(\nu)$, and $V_3 = N_{clusters}(\nu)$, the three usual MFs defined above plus an additional quantity called \textit{number of clusters}, $N_{clusters}(\nu)$, for $k = 3$. The latter quantity is the number of connected regions with height $\nu$ greater (or lower) than the threshold $\nu_t$ if it is positive (or negative), i.e., the number of hot (or cold) spots of the map. 

Consider a set of $m$ simulated CMB maps. For the $i$-th simulated map, with $i = 1, 2,..., m$, we compute the four MFs $\{V_k, \, k=0,1,2,3 \} \equiv (V_0,V_1,V_2,V_3)$ for $n$ dif\/ferent thresholds $\nu = \nu_{_{1}}, \nu_{_{2}}, ...\, \nu_n$, previously defined dividing the range $- \nu_{max}$ to $\nu_{max}$ in $n$ equal parts. Then, for the $i$-th map and for $k$-th MF we have not one element, but the vector 
\begin{eqnarray} \label{def_v}
\,\, v_k^i \equiv ( V_k(\nu_{_{1}}),V_k(\nu_{_{2}}), ... V_k(\nu_n) )|_{\mbox{for the $i$-th map}} \, .
\end{eqnarray}
In this work the values chosen for such variables are $\{\nu_{max},n\} = \{3.5,26\}$~\citep{2012/ducout}.

Once calculated the MFs vectors we define the training dataset, $T\{x_i,y_i\}$, for the NN, where $x_i$ is called input data and $y_i$ the output data, for $i = 1, 2, ..., m$, and $m$ is the number of simulated CMB maps. This training set configuration is necessary in order to allow the network to associate a certain kind of input pattern to a specific output. We set the $i-th$ input vector as the MF vector of the $i-th$ simulated map, $x_i = v_k^i$, for just the $k$-th MF. The output vector, $y_i$, is defined according to the number of classes $N_{class}$ of the input data. For example, if we use $N$ $classes$ of maps, corresponding to ensembles with dif\/ferent non-Gaussian levels, we have $N_{class} = N$. Then, our $m$ output vectors have $N$ elements, $y_i = (1,0,...,0)$ for $class = 1$, $y_i = (0,1,...,0)$ for $class = 2$, and so on, until $y_i = (0,0,...,1)$ for $class = N$. 

After this training process, the NN should be able to identify the same pattern in a dif\/ferent set of input vectors, e.g. the test dataset $\{x_j,y_j\}$, for $j= 1, 2, ..., l$. That is, providing to the already trained NN a dataset with $l$ input vectors $\{x_j = v_k^j\}$, it returns $l$ output vectors $\{y_j\}$.

At this point the estimator described here becomes dif\/erent from that one presented in our previous work. The \textit{classifier} estimator used the information contained in the output $\{y_j\}$ of the NN to classify each CMB map according to the class it belongs, that is, informing its non-Gaussian level ($f_{\rm \,NL}$ range). For the current analysis the NN's output are used to estimate the $f_{\rm \,NL}$ value for each MC CMB map and, afterwards, the Planck CMB maps. The details of this new approach will be given below.

\subsection{The NN's output as a $\widehat{f}_{\rm \,NL}$ estimator} \label{subsec:estimator_eq}

Still considering $N_{class} = N$, one can write the output vector of each $j$-map as being $y_j = \{y_j^c\} = (y_j^1, y_j^2, ..., y_j^N)$, where $y_j^c$ is expected to be $\sim 0$ or $\sim 1$. Using the $y^{c}_j$ elements, our estimator for the $f_{\rm \,NL}$ parameter is defined as

\begin{equation} \label{estimator}
\widehat{f}^{~j}_{\rm \,NL} \equiv \sum^{N_{class}}_{c=1} \langle f_{\rm \,NL} \rangle_c ~ y^c_j,
\end{equation}

\noindent 
where $\langle f_{\rm \,NL} \rangle_c$ is the mean of $f_{\rm \,NL}$ values considering that the primordial NG corresponds to the $c$-class of MC CMB maps. 

The initial idea to construct our estimator was inspired in the $f_{\rm \,NL}$ classification estimator defined by \cite{2011/casaponsa}, where the authors consider the classification problem using a NN in a probabilistic way. Their estimator is constructed such that each element of the NN's output vector is transformed into the probability of the corresponding input vector (from a CMB map) to belong to the $c$-class. 

After some analysis, we have checked that using the output vector in the exact form the NN returns it, that is, without transforming it into a probability quantity as in~\cite{2011/casaponsa}, our estimator works quite well. For this reason we decide to use this definition, Equation~\ref{estimator}. In the next section we expose --as an illustration-- the statistical analysis performed for one of the tests exposed here.

\section{Estimator application to synthetic and Planck data} \label{sec:application}

Aiming to improve the ef\/ficiency of our estimator in analysing Planck CMB maps we have trained and tested it in the more realistic (compatible) maps and analysed a fairly large number of dif\/ferent datasets.
These are composed by MC CMB maps contaminated with a mix of non-primordial non-Gaussian signals, in order to train more appropriately the NNs, that is, using input data more consistent with the Planck data. All performed analyses correspond to the three following classes: 
$f_{\rm \,NL} = [-20,20]$, \textit{class 1}; 
$f_{\rm \,NL} = [20,60]$, \textit{class 2}; and 
$f_{\rm \,NL} = [60,100]$, \textit{class 3}; 
using training and test datasets composed by $m/3=1900$ and $l/3 = 100$ perimeter-MF vectors, respectively, of each class of data\footnote{It is worth to mention here a curious and unexpected behaviour of the NN: when trained with a large dataset, say thousands of maps, all generated from the set of 2000 CMB base maps, or equivalently the set of 1000 coef\/ficients 
$\{a_{\ell \, m}\}$, just varying the $f_{\rm NL}$ values (equation \ref{alms_comb}), the NN looks like vitiated in the sense that it overestimates the efficiency in the classification of the tested maps. For this reason we decided to use only these, relatively small but {\em statistically independent}, 
training and test datasets.}. 

In addition, we also avoid possible discrepancies in the mapmaking parameters between synthetic and Planck maps. An example is the dif\/ference in the scale, since the Planck map was degraded to ${\rm \,N}_{\mbox{\small side}} = 512$ in order to have the same pixelization of the simulated maps, and another is the beam format of the instrumental beam, dif\/ferent from the Gaussian beam used to construct our simulations. In the case of the scale problem, differences can occur because MFs are sensitivity to the resolution scale of the temperature map. In Ref. \cite{2014/novaes}, it was observed a larger amplitude of the perimeter calculated from Planck map compared with those calculated from the synthetic ones. This problem is minimized performing a smoothing procedure on both maps (see Ref. \citep{2014/novaes} for details). Moreover, this procedure could even minimize ef\/fects of contamination by extragalactic sources, like SZ and lensing ef\/fects, since they contribute on small scales. For these reasons, we used a smoothing tool with a Gaussian beam of {\sc fwhm}$ = 10$ arc-min, chosen to be large enough to minimize any discrepancies, besides not interfering with the kind of analysis we are interested in, the constraints of local NG in large angular scale.

The next subsections present all performed tests, besides some statistical analysis and the results from the application of the trained NN to Planck CMB data.

\subsection{Tests on synthetic data: Ef\/fects of the Galactic residuals} \label{subsec:test_synthetic_data}

The main purpose of our analyses is to verify how much the estimator's performance is influenced by the presence of a mix of non-Gaussian contaminants in the synthetic data. For this, we performed several tests, checking the behaviour of the NN when trained upon 
dif\/ferent datasets, that is, MC CMB maps (on the above-mentioned classes) contaminated by a mixture of dif\/ferent types of Planck-like noise and, particularly important, we added weighted contributions of Galactic diffuse emissions (see all the combinations analysed in Table~\ref{tab:data}). 

Using each type of dataset to train and test NNs, we were able to use the Equation~\ref{estimator} to obtain the estimated $\widehat{f}_{\rm \,NL}$ values from the maps composing the test set. The results of these tests are summarised in Table \ref{tab:smica_result}, presenting results from the analysis upon synthetic maps contaminated by $\mathtt{SMICA}$-like noise, and Table~\ref{tab:se_ni_cr_results}, relative to the analysis upon synthetic maps contaminated by $\mathtt{SEVEM}$, $\mathtt{NILC}$ and $\mathtt{Commander-Ruler}$-like noise. The second and third columns of Table \ref{tab:smica_result}, and third of Table \ref{tab:se_ni_cr_results}, show the characteristics of the contamination by residual Galactic emission. Moreover, aiming to better understand the impact of the presence of the Galactic residuals to the ef\/ficiency of the estimator, we performed two additional tests: Test \#5, using the $\mathtt{SMICA}$ confidence mask instead of U73 which was applied to all the other tests; and Test \#6, where we include just the $\mathtt{SMICA}$-like noise as a non-primordial contamination to the CMB signal, without considering the foreground residuals.

\begin{figure}[h!]
\centering
\includegraphics[width=.7\linewidth]{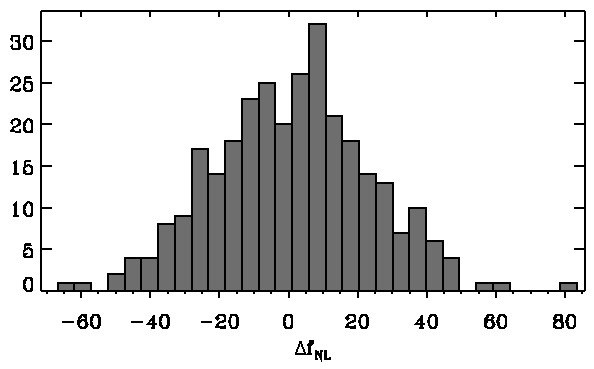}
\caption{Histogram of $\Delta f_{\rm \,NL}$ values obtained from Test \#4.}
\label{fig:hist_delta_fnl}
\end{figure}

In order to evaluate how well the $\widehat{f}_{\rm \,NL}$ values are obtained by the NN, we define the quantity
\begin{equation} \label{eq:delta_fnl}
\Delta f_{\rm \,NL} \equiv f_{\rm \,NL} - \widehat{f}_{\rm \,NL}.
\end{equation}
The columns 4-7 of Tables \ref{tab:smica_result} and \ref{tab:se_ni_cr_results} present the mean of the $\Delta f_{\rm \,NL}$ values for the three classes (i.e., $\langle\Delta f_{\rm \,NL}\rangle_c$, for the classes $c = 1, 2, 3$) and their corresponding standard deviations $\sigma(\Delta f_{\rm \,NL})$. Finally, the last column of both tables give us an idea of the maximum error attained by the NN, where we observe that $|\Delta f_{\rm \,NL}|^{\mbox{\tiny MAX}} \leq 88.9$ among all nine tests. 
To emphasize the role of the quantity defined in the Equation \ref{eq:delta_fnl}, we show in Figure \ref{fig:hist_delta_fnl} the histogram of the $\Delta f_{\rm \,NL}$ from, for example, Test \#4. 
It is possible to infer from this distribution that 95\% of the $\widehat{f}_{\rm \,NL}$ values are dif\/ferent from the input values by $|\Delta f_{\rm \,NL}|(95\%) \lesssim 44$, that is to say that the value $|\Delta f_{\rm NL}|^{\mbox{\tiny MAX}} = 78.5$ obtained in Test \#4, is just one isolated value as observed in the tail of the histogram of Figure~\ref{fig:hist_delta_fnl}. 

\vspace{0.5cm}

\begin{figure}[h!]
\centering
\includegraphics[width=.49\linewidth]{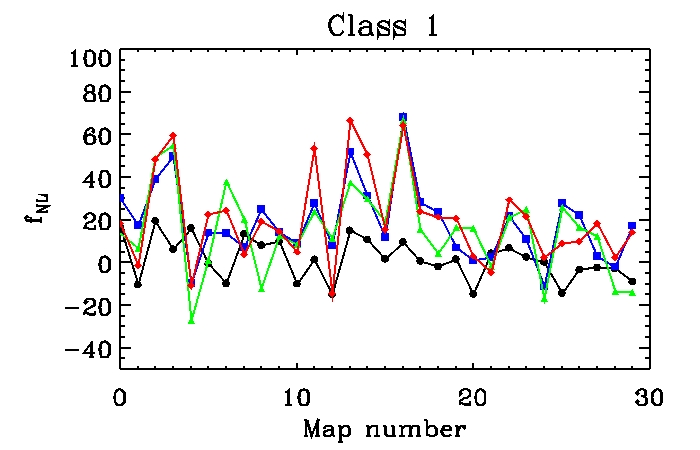}
\includegraphics[width=.48\linewidth]{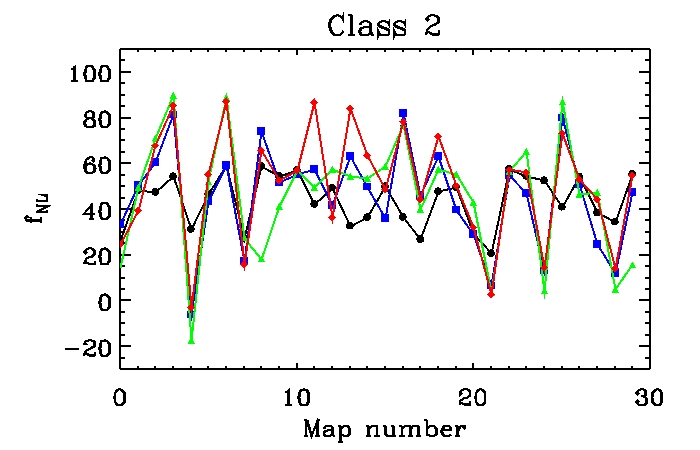}
\includegraphics[width=.48\linewidth]{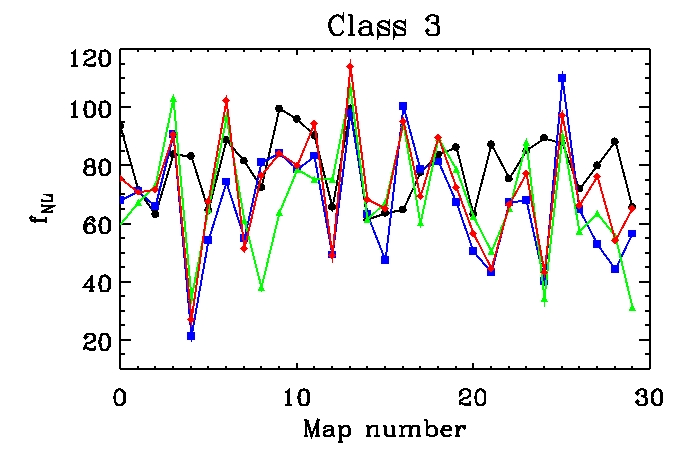}
\caption{Graphic of the input $f_{\rm \,NL}$ values (black) and the estimates, $\widehat{f}_{\rm \,NL}$, obtained from Tests \#4 (blue) and \#5 (green), both contamination by $\mathtt{SMICA}$-like noise and Galactic foreground residual, but using different masks, namely, U73 and $\mathtt{SMICA}$ confidence mask, respectively; and Test \#6 (red), contaminated only by $\mathtt{SMICA}$-like noise, with U73 cut-sky mask. We plotted the estimated values for 30 maps of each class.}
\label{fig:foreg_vs_noforeg}
\end{figure}

In addition, we can use the concept defined in Ref. \cite{2014/novaes}, to test the ef\/ficiency of our estimator in the current analyses. This is done counting the number of success (called \textit{hits}) of the NN in the classification of the vector $y$ corresponding to a map of the test set. This classification is based on the class to which it belongs, which is given by the largest of the three elements of the vector $y$. Summarising our results, in all tests here the number of \textit{hits} varies in the range 
$60\% - 70\%$, clearly less than the $\sim95\%$ of hits achieved in Ref. \cite{2014/novaes}, although obtained under dif\/ferent circumstances. This difference in the performances indicates that the current analyses with new features, like the continuity of the $f_{\rm NL}$ intervals defining the classes and, mainly, the smaller size of training datasets (see section~\ref{sec:application} for details), now restricted to the set of 2000 CMB base maps, contribute to diminish the ef\/ficiency of our estimator. From other side, observing Figure~\ref{fig:foreg_vs_noforeg} and also Table~\ref{tab:smica_result}, one can compare the results in the classification of the test maps, for each one of the three classes, for the Tests \#4, 5, and 6. Indeed, the estimates for these three tests show a great agreement between them, allowing to conclude that the addition of a mixture of non-primordial non-Gaussian signals contaminating the MC CMB test maps, and the use of a dif\/ferent mask, has no significant ef\/fect on the estimator performance.


\begin{table}[h]
	\centering
    \begin{tabular}{|c|ccccccc|}
    \hline
    Test & \multirow{2}{*}{Weight}      & Freq. & ~ & $\langle\Delta f_{\rm \,NL}\rangle_c$ & ~ & \multirow{2}{*}{$\sigma(\Delta f_{\rm \,NL})$} & \multirow{2}{*}{Max($|\Delta f_{\rm \,NL}|$)} \\
   \#       &                         & (GHz)   & $c = 1$ & $c = 2$ & $c = 3$ &       &              \\  \hline
    1       & \multirow{2}{*}{0.1 \%} & 70      & -15.6   & 0.3     & 18.8     & 25.7 & 84.7 \\
    2       &                         & 217     & -14.3   & -1.2    & 12.7     & 24.8 & 76.6 \\ \hline
    3       & \multirow{2}{*}{10 \%}  & 70      & -15.0   & -0.2    & 15.4     & 24.1 & 74.0 \\
    4       &                         & 217     & -16.3   &  1.7    & 17.4     & 22.8 & 78.5 \\ \hline
    5$^a$   &  10 \%                  & 217     & -10.4   &  1.9    & 15.9     & 23.2 & 88.9 \\
    6$^b$   &  ---                    &   ---   & -17.0   & -2.9    & 11.6     & 23.2 & 80.0 \\ \hline
\multicolumn{8}{p{13cm}}{$^a${\footnotesize The only test we use the corresponding $\mathtt{SMICA}$ confidence mask and not U73 mask. \newline $^b$ Non-primordial contamination given only by the $\mathtt{SMICA}$-like noise, i.e., without considering the foreground residuals.}}
    \end{tabular}
    \caption{Results of the tests on datasets contaminated by $\mathtt{SMICA}$-like noise.}
    \label{tab:smica_result}
\end{table}

\begin{table}[h]
\centering
\begin{tabular}{|c|ccccccc|}
\hline
Test  & \multirow{2}{*}{Noise-like} & Freq. &         & $\langle\Delta f_{\rm \,NL}\rangle_c$    &         & \multirow{2}{*}{$\sigma(\Delta f_{\rm \,NL})$} & Max \\
  \#  &                             & (GHz) & $c = 1$ & $c = 2$ & $c = 3$ &      & ($|\Delta f_{\rm \,NL}|$) \\ \hline
7     & $\mathtt{NILC}$             &  217  &   -19.2  &  -4.5   &  12.3    & 24.8 & 80.9  \\
8     & $\mathtt{SEVEM}$            &  217  &   -18.1  &  -2.7   &  11.3    & 23.3 & 80.4        \\
9     & $\mathtt{Commander-Ruler}$  &   70  &   -17.3  &  -2.4   &  14.7    & 23.9 & 71.8    \\ \hline
\end{tabular}
\caption{Results of the tests on datasets contaminated by $\mathtt{NILC}$-, $\mathtt{SEVEM}$-, and 
$\mathtt{Commander-Ruler}$-like noise.} 
\label{tab:se_ni_cr_results}
\end{table}

\begin{figure}[h!]
\centering
\includegraphics[width=.325\linewidth]{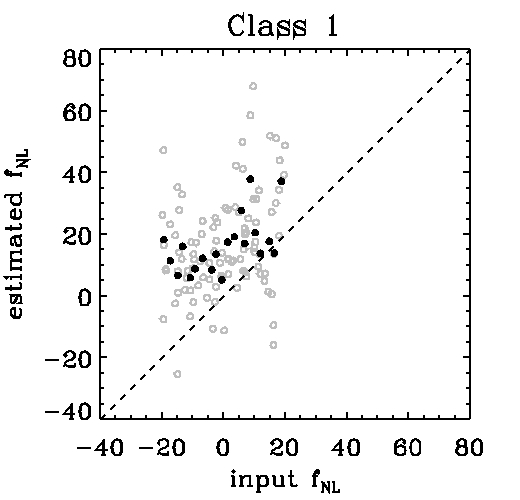}
\includegraphics[width=.325\linewidth]{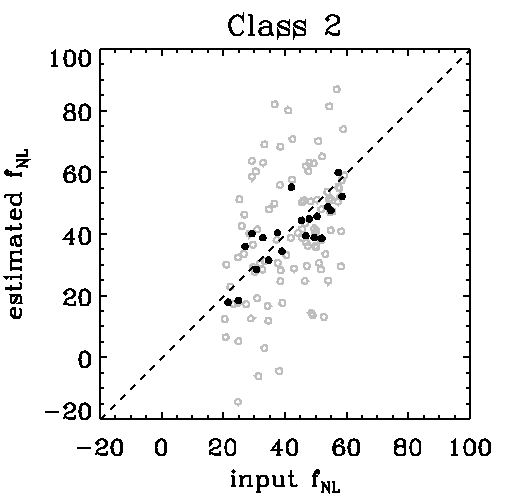}
\includegraphics[width=.325\linewidth]{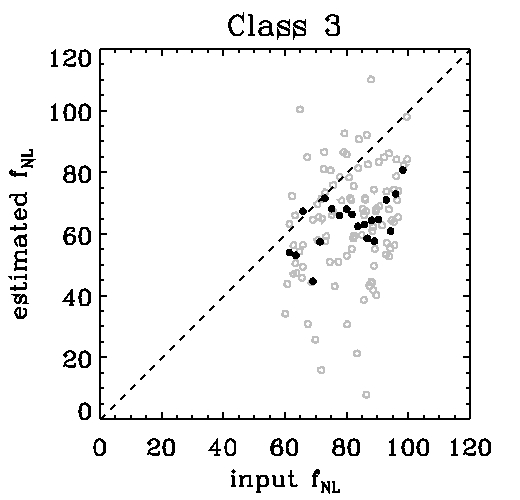}
\caption{Graph of estimated ($\widehat{f}_{\rm \,NL}$ from Test \#4) and input values of $f_{\rm \,NL}$ for each class (gray symbols). Black symbols correspond to binned intervals. The diagonal dashed straight is the equality line.}
\label{fig:fnl_test_vs_estim}
\end{figure}

The discussion above emphasises the good performance of the estimator, in addition to be a support to our choice in defining the estimator as in Equation \ref{estimator}, using directly the output vector for $\widehat{f}_{\rm \,NL}$ calculations. This choice can be even strengthened based on Figure~\ref{fig:fnl_test_vs_estim}, constructed using the results from Test \#4, which graphically exhibit the dispersion of the $\widehat{f}_{\rm \,NL}$ estimated values relatively to the input values. Directly comparing input and estimated values, besides also using binned intervals\footnote{The black symbols in Figure \ref{fig:fnl_test_vs_estim} corresponds to the binned representation of the relation between the $f_{\rm \,NL}$ and $\widehat{f}_{\rm \,NL}$ values.}, Figure \ref{fig:fnl_test_vs_estim} allows to verify the agreement between them: as clearly seen the black symbols follow the equality line fairly well. In addition, we can also observe from the binned representation and the $\langle\Delta f_{\rm \,NL}\rangle_c$ values in Tables \ref{tab:smica_result} and \ref{tab:se_ni_cr_results} that for lower values of $f_{\rm \,NL}$ the estimated values, $\widehat{f}_{\rm \,NL}$, are slightly shifted to higher values, and the opposite occurs for larger input values, $f_{\rm \,NL}$ (a similar behaviour was observed by \citep{2011/casaponsa}). However, this small bias is accounted in the error estimatives, $\sigma(\Delta f_{\rm \,NL})$.

\subsection{Applying a trained NN to Planck CMB maps}  \label{subsec:application_planck}

Previous section presented several statistical analyses showing the results from the application of the NN derived from Test \#4. We consider this NN as a good example because it is the most appropriate among the six NNs used to analyse the $\mathtt{SMICA}$ map. This choice is based on a crucial information (see section \ref{subsec:galaxy} for details): the type of residual contamination expected for the $\mathtt{SMICA}$ map, as discussed in~\cite{2014/planck-XII}, 
is an under-subtracted thermal dust emission. Since dust emission is more significant at high frequencies, we just chose the NN derived from simulated maps contaminated with Galactic signals in 217 GHz.

In this sense, it is important to emphasise that the application of our estimator in the analysis of the released Planck CMB maps has to be done using the most appropriately trained NNs, that is, those trained using perimeter-MF vectors derived from the most compatible simulations with real data. This is the reason why we have performed all above mentioned tests using those specific datasets for training our NNs. 
This allows us to conclude that the $\mathtt{SMICA}$, $\mathtt{NILC}$, $\mathtt{SEVEM}$ and $\mathtt{Commander-Ruler}$ CMB maps are more appropriately analysed using the NNs derived from Tests \#4, 7, 8, and 9, respectively. Notice also that, we based our choices considering the worse scenario, that is, the tests with large weight value, 10\%, in order to have a residual signal below a few $\mu$K in amplitude, as expected by the Planck Collaboration (see \cite{2014/planck-XII} for details).

The results from analysing the Planck CMB maps using the corresponding trained NNs are summarized in Table \ref{tab:fnl_planck_maps}, presenting the $\widehat{f}_{\rm \,NL}$ values and the associated error bars. We would like to emphasize the good agreement between our estimates and Planck and WMAP last results. Specially regarding the $\mathtt{SMICA}$ map, derived from the component separation method considered the leading method due to its good performance achieved on the FFP6 simulations, and, for this, the most extensively analysed CMB map by Planck collaboration~\cite{2014/planck-XII}.

\begin{table}[!h]
\centering
    \begin{tabular}{|c|cc|}
    \hline
    Test \# & Planck Map                 & $\widehat{f}_{\rm \,NL}$ \\ \hline
     4      & $\mathtt{SMICA}$           & 33 $\pm$ 23              \\
    13      & $\mathtt{NILC}$            & 38 $\pm$ 25              \\
    14      & $\mathtt{SEVEM}$           & 41 $\pm$ 23              \\
    15      & $\mathtt{Commander-Ruler}$ & 39 $\pm$ 24              \\ \hline
    \end{tabular}
\caption {Results of applying the trained NNs to the four foreground-cleaned Planck CMB maps. 
Error bars refer to the $1\sigma$ confidence level computed from the simulated data (see Tables \ref{tab:smica_result} and \ref{tab:se_ni_cr_results}).}
\label{tab:fnl_planck_maps}
\end{table}

\section{Conclusions and final remarks}  \label{sec:conclusion}

We employed here a combined MF and NN based estimator that reveals high performance in analyses aimed to constrain the primordial NG present in sets of simulated CMB maps, but considering the presence of weighted mixtures of residual Galactic foregrounds. In a previous work \cite{2014/novaes} it was presented a first version of this estimator, as well as several and exhaustive tests done to evaluate its performance in classifying synthetic datasets according to its $f_{\rm \,NL}$ range. Although results from its application to Planck data seemingly agreed with latest results from Planck and WMAP-9yr data analyses, the results were imprecise in constraining the $f_{\rm \,NL}$ value in Planck CMB maps.

In the present work we have upgraded our estimator in two main aspects: (1) improving the way how the output vectors of the NNs are treated, using their elements to directly estimate the $f_{\rm \,NL}$ values of the CMB maps (see Equation~\ref{estimator} for the definition of $\widehat{f}_{\rm \,NL}$), and (2) testing its performance in a wide range of likely and unlikely scenarios, and this was done performing the training processes upon diverse sets of simulated maps, contaminated with weighted residual Galactic contaminations besides inhomogeneous noise and primordial NG, constructed in such a way to achieve more realistic features, that, consequently, make them more compatible with Planck data. According to our results, the second point is the most important prescription to obtain a good performance of the estimator, what justifies all the performed tests. 

The tests were initiated aiming to evaluate the ef\/ficiency of the estimator when applied to simulated data contaminated by a mix of secondary non-Gaussian signals. Based on the statistical analyses of the obtained $\widehat{f}_{\rm \,NL}$ values, which were derived from Tests \#1 to 9 and whose results are presented in Tables \ref{tab:smica_result} and \ref{tab:se_ni_cr_results}, one can confirm the excellent performance of our estimator.

To emphasise the results achieved here observe that, as shown in Ref.~\cite{2014/novaes}, the performance of our combined estimator is related to the size of the training datasets. For this, it is expected that the efficiency of the NN in the current analyses could still be improved with larger sets of $\{a_{\ell \, m}^{G}\}$ and $\{a_{\ell \, m}^{NG}\}$, in order to produce a better training of the NN. Nevertheless, with the available set of synthetic maps for training the NNs, we still obtained fairly good values for the standard deviation $\sigma(\Delta f_{\rm \,NL})$, all in the range $22.8 - 25.7$ (calculated from the dif\/ference between the input and the estimated $\widehat{f}_{\rm \,NL}$ values). 

Using a methodological procedure we select a set of four NNs that we apply to the foreground-cleaned Planck CMB maps, obtaining the $\widehat{f}_{\rm \,NL}$ value in each case (see Table~\ref{tab:fnl_planck_maps}). The four NNs were trained using datasets simulated considering the same weight value (10\%) and obeying the informations given in~\cite{2014/planck-XII} regarding the expected residual contaminations by secondary non-Gaussian signals in Planck maps. For this reason, the dif\/ferences obtained between the estimates for each CMB map are comprehensive, and allow us to conclude that the Planck CMB maps could have a residual contamination higher 
than that ones considered in our simulations, or even other types of non-primordial signals. Therefore, from our final results, 
namely, $\widehat{f}_{\rm \,NL} = 38 \pm 25,~41 \pm 23$ and $~39 \pm 24$ for $\mathtt{NILC}$, $\mathtt{SEVEM}$, and $\mathtt{Commander-Ruler}$ respectively, and specially from the analyses of the $\mathtt{SMICA}$ map, with $\widehat{f}_{\rm \,NL} = 33 \,\pm\, 23$, corresponding to the $1\sigma$ confidence level, we conclude that our estimates are in excellent agreement with the latest constraints on this primordial local NG from WMAP-9yr~\cite{2012/WMAP9a} and Planck~\cite{2014/planck-XXIV} data analyses, i.e., $f_{\rm \,NL} = 38 \pm 18$, for the large angular scales. Ultimately, it is still worth to emphasize that all our analyses refer to the large angular scales where the ef\/fects of the primordial NG, with intensity $f_{\rm \,NL}$, seems to be more distinctive~\cite{2011/raeth,2013/modest}.


\acknowledgments

We are grateful for the use of the Legacy Archive for Microwave Background Data Analysis (LAMBDA) and of the $\{ a^{\mbox{\footnotesize G}}_{\ell \, m} \}$ and $\{ a^{\mbox{\footnotesize NG}}_{\ell \, m} \}$ simulations \citep{2009/elsner}. We also acknowledge the use of CAMB (http://lambda.gsfc.nasa.gov/toolbox/tb\_camb\_form.cfm), developed by A. Lewis and A. Challinor (http://camb.info/), and of the code for calculating the MFs, from \cite{2012/ducout} and \cite{2012/gay}. Some of the results in this paper have been derived using the HEALPix 
package~\citep{2005/gorski}. AB acknowledges a CAPES PVE project, and CPN acknowledges a Capes fellowship. CAW acknowledges the CNPq grant 308202/2010-4. We would like to thank the anonymous referee for very useful comments and feedback on this paper.



\end{document}